# In-situ identification of various structural features of vanadyl porphyrins in crude oil by high-field (3.4 T) ENDOR spectroscopy combined with DFT calculations

- Timur Biktagirov[a,b], Marat Gafurov[a,*], Georgy Mamin[a], Irina Gracheva[a], Andrey Galukhin[b,c], Sergei Orlinskii[a]

[a] Institute of Physics, Kazan Federal University, 420008 Kazan, Russia
[b] Institute of Geology and Petroleum Technology, Kazan Federal University, 420008 Kazan, Russia
[c] Institute of Chemistry, Kazan Federal University, 420008 Kazan, Russia

*e-mail: marat.gafurov@kpfu.ru

**Abstract**

Structural characterization of metalloporphyrins in complex systems such as native hydrocarbons is in the focus of scientific and industrial interests since many years. We describe electron-nuclear double resonance (ENDOR) of crude oil from the well without any additional sample treatment (i.e., in the native environment) in the magnetic field of about 3.4 T and temperature of 50 K by applying microwave pulses at 94 GHz (W-band) and radiofrequency pulses at near the proton Larmor frequencies of 144 MHz to probe the paramagnetic vanadyls. By means of density functional theory (DFT) calculations, ENDOR features are explained and ascribed to certain vanadyl porhyrin structural forms known to be present in crude oil.

## 1. Introduction

Metalloporphiryns in heavy oils were first discovered in 1930s by Treibs [1, 2], who postulated the biological origin of these compounds. Specifically, the porphyrins are formed from biological chlorophyll, bacteriochlorophylls, hemes, and other tetrapyrrolic biochemicals. Vanadyl ($VO^{2+}$) porphyrin is one of the most abundant forms of metalloporphyrin, and is of particular significance as a biomarker in the context of oil formation research [3]. Their chromatographic profiles provide information on the deposition environment [4-6]. That is why structural characterization of these vanadyl compounds has been addressed extensively in publications during the last decades. These studies appear even more interesting in light of the new exciting results proposing vanadyl complexes as a part of quantum computers [7, 8].

In general, petroleum porphyrins exist in homologous manifolds of several structural classes and can manifest great structural diversity [9, 10]. The molecule can contain different types of substitutions, including alkyl, cycloalkane, and aromatic groups. Among the possible structures of vanadyl porphyrins identified in natural hydrocarbons the following forms were considered to be the most common ones: etioporphyrin (VOEtio), deoxophylloerythroetioporphyrin (VODPEP) and benzoetioporphyrin (VOBenzo) [11].

There is no exclusive procedure for the isolation of porphyrin complexes from the host material [12]. Presently, one of the most promising and well-established protocols for structural characterization of metalloporphyrins involves the application of Fourier transform ion cyclotron resonance mass spectrometry [9, 10, 13, 14]. Due to the relatively low concentration of vanadium compounds and the very complex chemical composition of heavy crude oil, the successful identification of porphyrins requires chromatographic separation or demetalation. Nevertheless, the examples of successful direct observation of vanadyl porphyrins in unfractionated asphaltenes were reported in literature (as for the first time, by Qian et al. [14]).

An increased interest for studying unfractionated hydrocarbons by electron paramagnetic resonance (EPR) and related methods is to obtain last decade (see [15-18] and references therein). EPR based techniques do not face the problem of extraction or resolving vanadyl-porhyrins from the host matrix background, since it specifically detects vanadium compounds, probes their structure and dynamical properties. Besides of the higher sensitivity and spectral resolution, the advantages of the high-field EPR approaches for the identification of different paramagnetic complexes and their characterization in native oil containing formations are nicely reviewed and shown in [17]. For instance, in the case of even slight anisotropy of the magnetic interaction tensors (such as $g$- and hyperfine, $A$, tensors), high-field EPR can provide the information about their orientation dependencies and thereby about the spatial structure of the paramagnetic complex in disordered systems.

Some particular utilizations of the pulsed high-field EPR by means of the electronic relaxation times measurements for the establishment of the nature of crude oils and structures of the "free" radical – vanadyl paramagnetic complexes were demonstrated by our group [18, 19]. We have to acknowledge the papers [20-22] that prompt us to use double resonance techniques for the crude oil investigations. The feasibility of pulsed EPR for the vanadyls characterization in oil asphaltenes was shown in [23].

Though hydrocarbon systems contain sufficient amount of paramagnetic centers (up to one paramagnetic centre per asphaltene molecule) and different nuclei with the magnetic moments (see ref. [15-19] and literature cited there), surprisingly little is done in the field of electron-nuclear double resonance (ENDOR) studies of oil containing materials, their solutions and fractions at least as we can judge from the open sources. The most of the mentioned research are done on the model systems – "pure" vanadyls, diluted oil fractions, etc. Complex nature of native hydrocarbons, which consist from the mix of different vanadyls, the presence of the organic "free" radicals, which nature in oils and bitumen is still undefined, other paramagnetic species like Mn(II) require application of some special approaches and tools for their investigations by ENDOR which are not established and need to be further developed.

Herein, we present the first application of pulsed high-field (in the magnetic fields of about 3.4 T that corresponds to 94 GHz of the microwave frequency) ENDOR for studying the structure of vanadyl porphyrins in a heavy crude oil from the Russian oilfields without any additional treatment of the sample. As well, we use the advantages of modern computational chemistry in order to rationalize the observed spectroscopic features by means of density functional theory (DFT) calculations. The capabilities of this combined approach were recently discussed for the case of the model vanadyl porphyrin VOTPP in silica studied by 9.5 GHz ENDOR [24]. As concerning our group, we have exploited the combined high-field ENDOR-DFT approach for the investigations of structure and positions of some impurities in hydroxyapatite based materials [25, 26]. Quite recently, we have managed to observe complex pulsed ENDOR spectra caused by $^{14}$N nuclei at 9.5 GHz in the untreated heavy crude oil sample. The results with the corresponding DFT calculations for exact determination of EPR parameters including the main values and mutual orientations of $g$-, $A$-, and electric quadrupole, $P$, tensors are presented in ref. [27].

In the present work we explore $^{1}$H ENDOR also because these nuclei are situated on the sides of the complex and thereby have a potential to track the environmental changes (in contrast with $^{14}$N and $^{15}$N nuclei which are "buried" inside the complex [7, 8, 27]). Since ENDOR method is sensitive to the precise positions of the surrounding atoms around the metal ion, we believe that it is capable to complement the existing analytical approaches by providing the unique structural data about vanadyl porphyrins even in complex organic mixtures.

## 2. Materials and methods
### 2.1. Sample preparation

Different oil and bitumen species from Russian oilfields were investigated by ENDOR. The results are shown for heavy crude oil sample from Mordovo-Karmalskoye oilfield of Volga-Ural basin, Republic of Tatarstan and light one from the West-Siberian basin (Table 1). They were used as received, without any additional sample treatment. The samples were put into the 0.4 mm inner diameter quartz capillary, then sealed and placed into the EPR resonator. Small volume of the species for the W-band measurements (of about 500 nl) can be regarded as an additional advantage of the high-field EPR – the major part of the material could be examined by other analytical methods, be subject of different kinds of manipulations, etc.

**Table 1.** Samples, their physical properties at room temperature (20 $^{o}$C) and the results of their SARA (saturated fractions, aromatics, resins and asphaltenes) analysis

|  | origin | density, (kg/m$^3$) / API gravity | viscosity, (mPa·s) | saturates and aromatics (%) | resins (%) | asphaltenes (%) |
|---|---|---|---|---|---|---|
| Sample 1 | Volga-Ural Basin | 945 / 18.2 | 1020 | 65.0 | 31.0 | 3.9 |
| Sample 2 | West-Siberian Basin | 837 / 36.6 | 6.0 | 67.7 | 8.3 | 0.45 |

### 2.2. EPR, ENDOR measurements

Pulsed EPR measurements were done using W-band (microwave frequency of about 94 GHz) Bruker Elexsys 680 spectrometer equipped with the flow liquid helium cryostat. Electron spin echo (ESE) EPR spectra were recorded by means of field-swept (by changing the induction of the external magnetic field $B_0$) two-pulse echo sequence $\pi/2 - \tau - \pi$ with the pulse length of $\pi$ pulse of 36 ns (W-band), and time delay $\tau = 240$ ns.

Pulsed ENDOR spectra were detected by Mims sequence $\pi/2$-$\tau$-$\pi/2$-$T$-$\pi/2$ with an additional radiofrequency (RF) pulse $\pi_{RF} = 16$ μs inserted between the second and third microwave $\pi/2$ pulses (Figure 1). RF frequency could be swept in the range of (1-200) MHz.

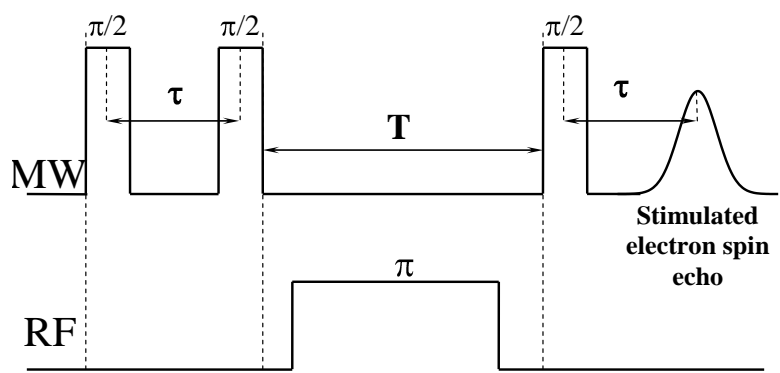

**Figure 1.** Mims pulse sequence at microwave and radiofrequencies used to obtain the ENDOR spectra as a function of stimulated electron spin echo amplitude from the frequency of RF pulse. The values of time delays ($\tau$, T) and pulse lengths ($\pi/2$ for MW and $\pi$ for RF) are given in the text.

Because the high-field EPR and ENDOR (particularly in the pulsed mode) are not so common for the petrophysical and chemical laboratories, the features of the $VO^{2+}$ EPR and some basics of ENDOR essential for the interpretation of the observed spectra are briefly described below.

The largest interaction in the high-field experiments is the Zeeman one between the magnetic moment of the unpaired electronic spin $S$ and $B_0$. The $VO^{2+}$ paramagnetic complex is characterized by $S = ½$ and, therefore, by two projections of $m_S = ±1/2$ onto the direction of $B_0$. The electronic Zeeman interaction is described by the parameter called $g$-factor with the typical values close to 2.0. Another interaction (common applicable to hydrocarbon systems) is the spin-nuclear one. As a nucleus could also have a magnetic moment, the interaction of the unpaired electron with the nucleus splits the electron energy levels, generating a structure called hyperfine structure in the EPR spectrum – each energy level splits into a closely spaced group of *(2I+1)* sublevels (here *I* is the nuclear spin quantum number), described by the hyperfine parameter *A*. For $^{51}V$ nuclear spin $I = 7/2$. For the calculations of energy levels of paramagnetic complexes in high magnetic fields one should also not forget the nuclear Zeeman interaction between *I* and $B_0$.

Taking into account that $VO^{2+}$ has a near planar symmetry, the $g$- and $A$-parameters should be regarded as anisotropic parameters having axial symmetry, i.e, $(g, A)_∥$ and $(g, A)_⊥$ (see also Section 3). Figure 2 presents the calculated electronic-nuclear energy levels of $VO^{2+}$ paramagnetic complex, EPR transitions and the corresponding EPR absorption line with the values of $(g, A)_∥$ (i.e., perpendicular to $VO^{2+}$ plane, $z$ axis) and $(g, A)_⊥$ (i.e., in the $VO^{2+}$ plane or *xy* plane) taken from [28] which will be appeared close to the obtained in the experiment (see Section 3). As for the most $VO^{2+}$ related investigations, here and below we suppose that the direction of the main axes coincide for $g$- and $A$- tensors. The powder EPR spectrum of $VO^{2+}$ represents the 2×8 hyperfine patterns (the projection of *I* is allowed to take 8 values: of $m_I$: ±7/2; ±5/2; ±3/2; ±1/2) for the parallel and the perpendicular complex orientations [15-18]. Only transitions are allowed with $\Delta m_S = ±1$ and $\Delta m_I = 0$.

For the orientation selected experiments it is quite convenient and usual to choose such values of $B_0$ that correspond to "pure" parallel and perpendicular orientations. It is not a trivial task for such complex systems as oil and bitumen (see Section 3). In this work we have chosen the values that correspond mainly to the $z$ axis perpendicular ($B_{1'}$) and parallel ($B_{2'}$) to the direction of magnetic field for $m_I = 3/2$ transition due to the next reasons: (1) the sufficient ESE amplitudes to obtain reasonable signal-to-noise ratio for the appropriate time for both values/orientations; (2) absence of overlapping with the shoulders of the "free" radical signal (see Section 3 and our papers [18, 19]); (3) absence of the orthogonal contributions (see Figure S1 in SI); (4) the using of the experimental data from the same value of $m_I$ simplifies the following interpretation and theoretical analyses. In refs. [20, 24] the values of $B_0$ corresponding to $m_I = -7/2$ (for parallel) and $m_I = -3/2$ (for perpendicular) orientations in the X-band for the conventional ENDOR experiments were chosen. Obviously, in the W-band these transitions ($m_I = -7/2$ and and $m_I = -3/2$) would be not our first choice (cf. Figure 2 and Figure 3).

In the fields $B_1$ and $B_2$ (Figure 3) the ENDOR signals were acquired by using Mims pulse sequence (Figure 1).

In the case of a "free" nucleus, RF pulse applied at the Larmor frequency

$$\nu_{Larmor} = |\gamma B_0| \equiv h^{-1}|g^{(I)}\beta^{(I)}B_0|, \qquad (1)$$

where $\gamma$ is a gyromagnetic ratio of the nuclear spin *I*, *h* is a Planck constant, $g^{(I)}$ is a nuclear g-factor and $\beta^{(I)}$ is a nuclear Bohr magneton, can change the state of the nuclear spin (the population of the nuclear sublevels). For proton ($^1H$) with $I = ½$, $\gamma_{1H} = 42.576$ MHz/T that results in $\nu_{Larmor} \approx 144.76$ MHz for $B_0 = 3.4$ T.

In the case of a coupled electron spin $S$ and a single nuclear spin such changing can modify the state of the electron spin (can change the population of the energy levels contributing to the EPR spectrum). For the hyperfine coupling constant $A$ and simple electron-nuclear coupling ($S = 1/2$, $I = 1/2$), it can lead to the appearance of the characteristic features in the ENDOR spectrum at the RF frequencies

$$\nu_{ENDOR} = \nu_{Larmor} \pm A/2 \text{ or} \quad (2)$$
$$\nu_{ENDOR} = A/2 \pm \nu_{Larmor}, \quad (3)$$

depending on the ratio between $A$ and $\nu_{Larmor}$. In the case of more than one nucleus coupled with electron spin, one can observe the multiple signals around the corresponding Larmor frequencies or $A/2$ with different hyperfine coupling constants or $\nu_{Larmor}$.

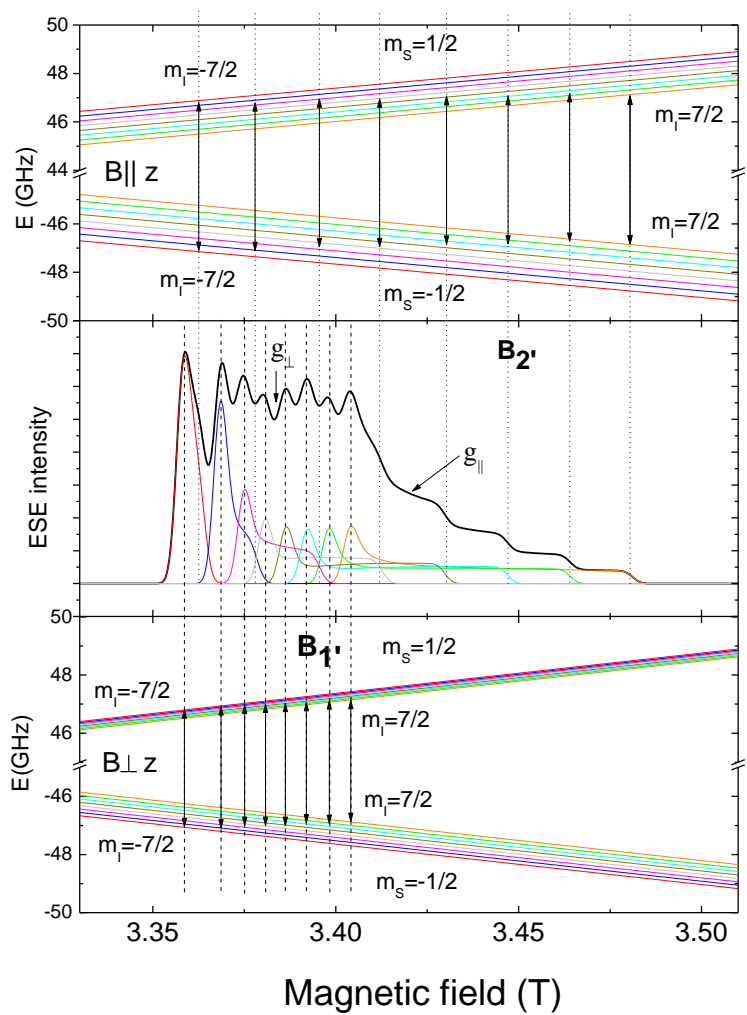

**Figure 2.** The energy levels (the top panel for $B_0 \parallel z$ while the bottom one for $B_0 \perp z$) and the corresponding absorption EPR spectrum (central panel) for $VO^{2+}$ complex calculated for the microwave frequency $\nu = 94$ GHz, $g_\parallel = 1.961$, $g_\perp = 1.984$, $A_\parallel = 470$ MHz, $A_\perp = 150$ MHz. Particular contributions from every EPR transition are color marked. Calculations are done in EasySpin package for Matlab [29] in the assumption that the main axes of $g$- and $A$- tensors coincide. The arrows mark the values of the magnetic field equivalent to the values of $g_\parallel$ and $g_\perp$. The values of $B_0$ that correspond to the EPR transitions with $m_I = +3/2$ are denoted as $B_{1'}$ and $B_{2'}$.

The ENDOR splitting $a_{ENDOR}$ can help not only to identify a type of nuclei coupled with the electron spins but also provide spatial relationships between them. For the pure electron-nuclei dipole-dipole interaction in the point model, the electron-nuclear distance, $r$, from the ENDOR splitting can be estimated from:

$$a_{ENDOR} \propto g \cdot g^{(I)} (1-3\cos^2\Theta)/r^3, \qquad (4)$$

where $g$ and $g^{(I)}$ are g-factors of electron spin $S$ and nuclear spin $I$, respectively, $\Theta$ is an angle between the directions of the electron-nuclear distance vector and $B_0$. As it follows from (4), $a_{ENDOR}$ depends on the distance between the electron and nuclear spins and their mutual orientation. In this work we consider ENDOR splittings due to the interaction of $VO^{2+}$ electron spin and $^1H$ nuclear spins with $S = I = ½$. One can find more details concerning different aspects of ENDOR technique and applications in ref. [26, 27, 30].

### 2.3. Computational details

All quantum chemical calculations were performed at the DFT level of theory with Orca software version 3.0 [31] using hybrid PBE0 exchange correlation functional [32]. Various structural models of vanadyl porphyrin with different combinations of side groups (including VOEtio, VODPEP, VOBenzo) as well as a bare porphyrin skeleton (denoted as VO throughout the following text) were considered. Geometry optimization of vanadyl complexes was carried out with 6-31G* basis set used for all atoms. For the computation of g-tensors and hyperfine coupling parameters, the 'Core prop' (CP(PPP)) basis set was applied for vanadium, and EPR-II basis set [33] was exploited for all other atoms. The calculated hyperfine interaction parameters for $^{51}V$ nucleus included the second-order term that arises from spin-orbit coupling. The size of the integration grid was increased for vanadium atom to minimize numerical errors. The EPR and ENDOR spectra were simulated using EasySpin toolbox for Matlab [29]; standard ENDOR procedure "salt()" was used (see Appendix S1) with orientation selection assured by choosing an appropriate excitation window (cf. Figure S1).

### 3. Results and Discussion

Figure 3A presents the field-swept pulsed EPR spectra of heavy oil sample #1 detected at 94.6 GHz at room temperature. The simulation of the spectrum displays the presence of two paramagnetic centers. The first one can be identified as vanadyl porphyrin complex with axial symmetry (cf. Figure 2) and the spectroscopic parameters listed in Table 2. The obtained experimental and calculated data are in a good correspondence with those for different vanadyls model systems [24, 28, 34]. The second signal with g-factor of 2.004 is attributed to stable organic "free" radical(s) [15, 35]. The EPR spectrum for sample #2 is very similar to the described above and is not shown. We notice that in the high-field experiment these signals are spectrally well resolved allowing suppressing the influence of "free" radicals on the measured vanadyls EPR and ENDOR spectra (cf. with refs. [18, 27]).

**Table 2**. Experimental principal values of g-tensor and $^{51}V$ hyperfine coupling tensor A (in MHz) of vanadyl porhyrin in crude oil compared with the DFT calculated values for VO molecule[*]

|      | $g_X (g_{max})$ | $g_Y (g_{mid})$ | $g_Z (g_{min})$ | $A_X$ | $A_Y$ | $A_Z$ |
|------|-----------------|-----------------|-----------------|-------|-------|-------|
| Exp. | 1.9845 | 1.9845 | 1.9640 | \|156.9\| | \|156.9\| | \|470.8\| |
| DFT  | 1.9867 | 1.9867 | 1.9710 | -157.0 | -157.0 | -463.9 |

[*]Estimated errors in experimental spectroscopic parameters are within $1\cdot10^{-4}$ for g-values and 5 % for hyperfine couplings constant. As follows from the calculations, the orientation of $g_Z$ and $A_Z$ axes coincides with the molecular symmetry axis.

No essential changes of the W-band spectra were obtained at the lower temperatures. We have chosen the value of T = 50 K because at this temperature the electronic relaxation times of vanadyl complexes were obtained to be optimal to acquire the ENDOR spectra in all samples with the sufficient signal-to-noise ratio for the reasonable measurement time (of about 2 hours).

The W-band $^1$H ENDOR spectra measured at $B_1$ and $B_2$ are shown in Figure 3B. As Eq. (4) shows, the values of these splittings depend on the magnitude of the hyperfine interaction and the relative orientation of the electron and nuclear spins with respect to external magnetic field.

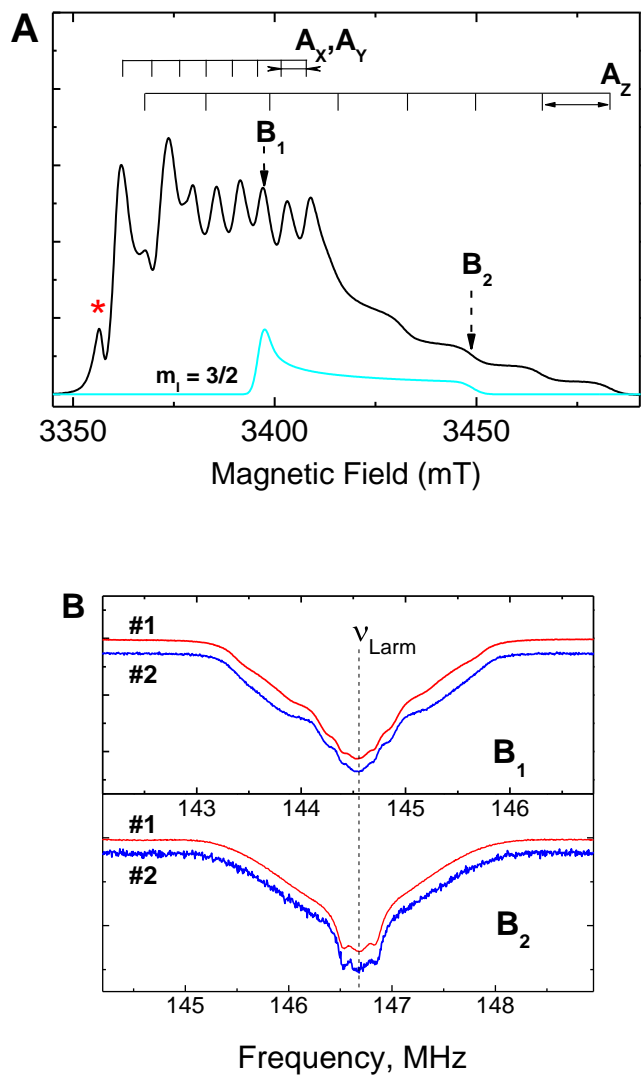

**Figure 3.** (A) W-band EPR spectrum of vanadyl porphyrin in crude oil sample # 1 in pulse mode at T = 50 K and repetition time of 0.5 ms (repetition rate of 2 kHz). The curve below shows the simulation of a hyperfine transition with $m_I = +3/2$, for which the measured ENDOR spectra are presented (with $B_1$ and $B_2$ being the corresponding magnetic fields related to $g_Z$ axis parallel and perpendicular to the direction of magnetic field; cf. Figure 2). The signal with g-factor of 2.004 assigned to organic radicals is marked by asterisk. (B) $^1$H Mims ENDOR spectra corresponding to different molecular orientations of vanadyl porphyrin detected in the vicinity of proton Larmor frequency at T = 50 K for the crude oil samples #1 and #2.

In general, one can expect the $^1$H ENDOR spectra to contain the contributions from the statistically averaged ensemble of protons. This involves both the protons contained in the porphyrin itself and those of the surrounding molecules. If the distribution of the protons around vanadyl ion was purely random, one would observe a single broad spectral line centered at the proton Larmor frequency. On the other hand, the protons bound to the porphyrin molecule should give rise to the more distinctive spectroscopic features because of statistical dominance (cf. the splittings in Figure 3B). Therefore, from the pattern of hyperfine splittings it is possible to infer about the chemical composition of vanadyl porphyrin molecules in the studied samples. With that being said, we aim to attribute the resolved spectroscopic features to the particular positions of the protons in the porphyrin molecule.

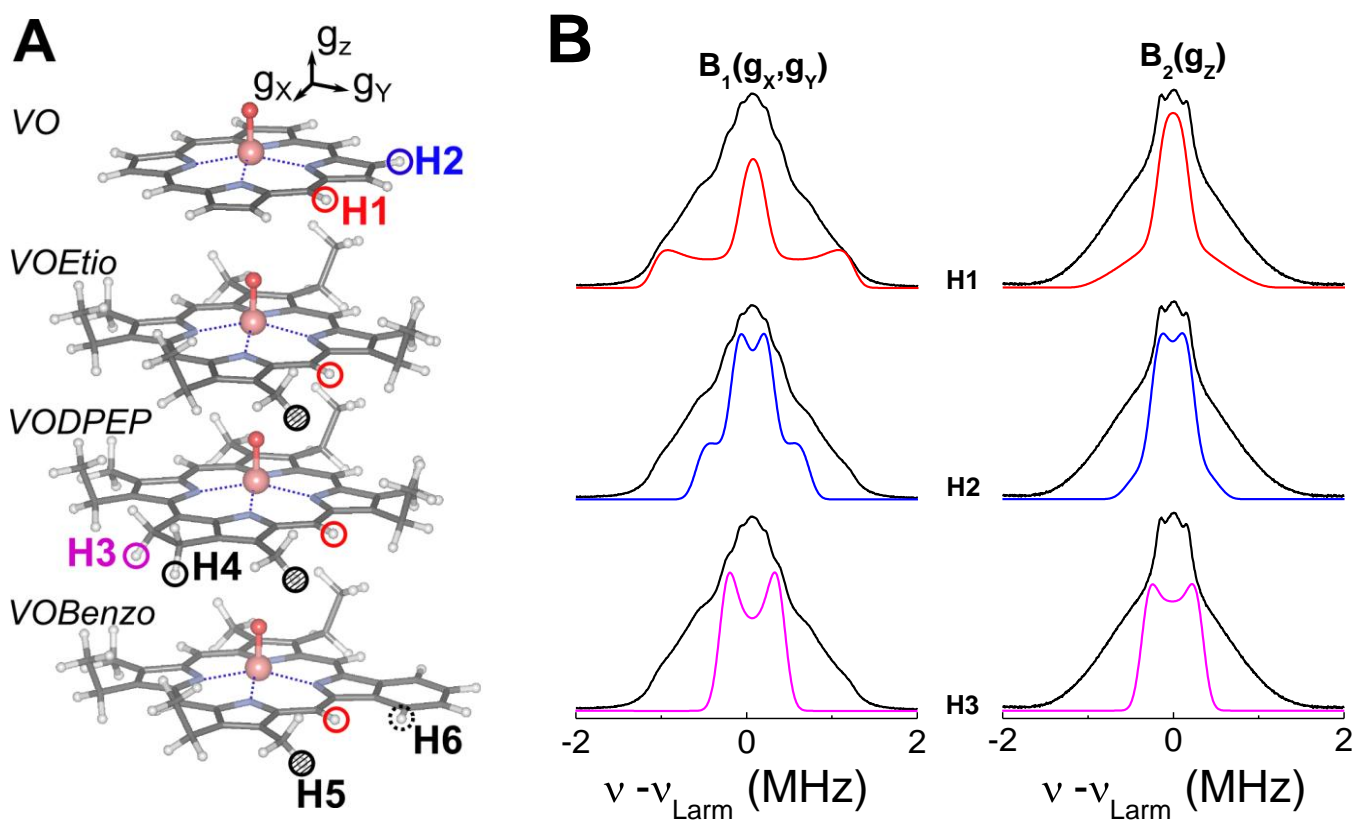

**Figure 4.** (A) Optimized chemical structures of vanadyl porphyrin models (VO, VOEtio, VODPEP, VOBenzo [11]). Circles indicate the positions of the representative protons of the porphyrin skeleton (H1 and H2) and those attributed to the possible classes of side groups (H3-H6). The illustrated orientation of the g-tensor corresponds to VO molecule. (B) ENDOR spectra simulated (red, blue, and magenta) for the selected protons presented in comparison with the experimental spectra (black). The intensities of the simulated spectra are scaled to be less than for the experimental ones.

In order to do this, we calculate $^1$H hyperfine coupling parameters for different possible structures of the vanadyl porphyrin within the framework of DFT. First of all, we consider the most trivial VO model that contains only two types of protons: four protons bound to the methine links between pyrrole groups (H1) and eight pyrrole protons (H2). The calculated g-values and $^{51}$V hyperfine coupling parameters for this model are listed in Table 2 and show a reasonable agreement with the experimental values. The hyperfine coupling parameters derived for the H1 and H2 protons are presented in Table 2. It should be noted that due to the axial symmetry of the paramagnetic center, the protons of the same type give almost the identical contributions to ENDOR spectrum.

In a real porphyrin molecule, both H1 and H2 can be substituted. Thus, we consider the standard structural models of natural vanadyl porphyrins, namely VOEtio, VODPEP and VOBenzo forms [11]. Furthermore, we perform additional calculations for the structures comprised of different possible classes of the substituting side groups (such as alkyl, cycloalkane, aromatic) [9, 10]). We report that all these structures manifest only minor differences in calculated g- and $^{51}$V hyperfine tensors (data are not shown). For each considered structural model, the tensors of $^{1}$H hyperfine coupling were calculated. Notably, we find that the addition (or removal) of the different side groups for the most cases does not affect the hyperfine coupling parameters related to the other nearby protons. The principle values obtained for the possible protons closest to the vanadium ion (thereby giving the largest hyperfine splittings) are listed in Table 2.

**Table 2**. DFT calculated $^{1}$H hyperfine coupling parameters (in MHz), interatomic distances (in Å) for the selected porphyrin protons attributed to the possible structural models (cf **Figure** 4) and the corresponding Euler angles (in degrees)*

|     | $A_X$ | $A_Y$ | $A_Z$ | $r_{V-H}$ | α | β | γ |
|-----|-------|-------|-------|-----------|-----|-----|------|
| H1  | -0.2  | -0.4  | 2.4   | 4.6       | 89.9 | 84.1 | -175.3 |
| H2  | -0.4  | -0.4  | 1.3   | 5.3       | 99.2 | 84.1 | -55.7  |
| H3  | -0.6  | -0.6  | 0.9   | 5.6       | 83.9 | 106.8 | -176.3 |
| H4  | -0.3  | -0.3  | 0.8   | 6.0       | 75.3 | 74.2 | -19.9  |
| H5  | -0.4  | -0.4  | 0.8   | 6.0       | 81.6 | 73.1 | -63.3  |
| H6  | -0.4  | -0.4  | 0.7   | 6.0       | 86.6 | 83.0 | -106.5 |

*Estimated errors of the calculated hyperfine parameters are within 20% (based on the observed discrepancies between the values obtained for each type of proton in different structural modifications and different sites of the molecule). Euler angles (α β γ) are presented for a selected $^{1}$H nucleus and specify *Z-Y-Z* rotation that transforms the molecular frame with the *Z*-axis being perpendicular to the porphyrin plane (see Figure 4) to the frames where the hyperfine tensors for $^{1}$H are diagonal.

As Figure 4B illustrates, the pattern of hyperfine splittings and its change with respect to the molecular orientation can be reasonably reproduced by a set of the three types of the closest porphyrin protons (H1, H2, and H3). One can though argue that the contribution of H2 is somewhat doubtful, since in the majority of the proposed structural modifications of natural vanadyl porphyrins this proton is implied to be substituted by side groups [9-11]. Nevertheless, to our knowledge these are the only possible candidates to give such high values of hyperfine coupling constants. The protons situated at larger distances, i.e. on cycloalkane (H4), methyl (H5) and aromatic (H6) side groups, only contribute to the region ± 0.4 MHz near the Larmor frequency. Thus, the provided analysis proves the feasibility of ENDOR for identification of the particular structural forms of vanadyl porphyrins in the oil-containing systems.

At the same time, the external protons from axial ligands could manifest rather large values of hyperfine coupling [24]. However, in a complex system (such as crude oil samples studied in this work) these protons are not expected to give rise to narrow and distinctive spectroscopic splittings due to statistical averaging. Instead, those would contribute to the ENDOR spectrum as a broad background. The presence of such contributions from intermolecular hyperfine interactions in turn can explain the fact that the sum of the three closest porphyrin protons would not reproduce the overall shape of the experimental spectra. This observation leads to a conclusion that ENDOR spectroscopy has a potential to shed light on the mechanisms of aggregation of vanadyl porphyrins with other organic molecules. This can be extremely useful for studying crude oils under the catalytic or thermal treatment, where ENDOR can successfully complement the established capabilities of conventional EPR [15, 16, 35].

In addition, in the classical ENDOR work [34] the dependence of $^1$H spectra of vanadyl complex VOTPP on the used solvent was pointed out. It again emphasizes the importance of investigation of vanadyls behavior in native environment and gives an additional motivation to track the EPR and ENDOR changes with the oil samples treatments.

We have to note that in general Mims-type ENDOR suffers from the so-called "blind spot" behavior [30, 36]. ENDOR spectra recorded with a particular time delay τ' are not sensitive to the frequencies $\nu_B$ which are defined as $|\nu_B - \nu_{Larmor}| = n/(2\tau')$, where $n = 0, 1, 2...$ The "blind spots" could lead to the distortion of ENDOR spectra as line intensities around the "blind spots" could be considerably attenuated. As it is seen, in our experiments these "blind spots" should appear out of the $^1$H ENDOR spectra we report or on the spectra "shoulders" and no signal drop in the vicinity of $\nu_{Larmor}$ is to obtain (Figure 3B). Therefore, the presented in this manuscript data do not suffer from this effect although for the potential application of the described approach for the quantitative estimations of the relative contributions of different types of vanadyls should be certainly taken into account.

## 4. Conclusions

In this work we report observation of the reproducible pattern of hyperfine splitting in the $^1$H ENDOR spectra of vanadyl for two different crude oil samples. From the analysis of ENDOR spectra combined with DFT calculations, it was possible to ascribe the observed spectroscopic features to the particular structural modifications of porphyrin molecule known to be present in crude oil [11, 28]. Thus, we conclude that $^1$H ENDOR allows resolving the certain structural forms of vanadyl porphyrin in a complex hydrocarbon mixture. In this sense, we believe that pulsed ENDOR method can complement mass-spectrometry and become a part of a standard analytical toolkit. Since it is a non-destructive method which requires a small amount of material (less than one microliter) and does not need a special sample preparation procedure, it can be very useful for preliminary characterization of crude oils prior to chromatographic separation.

Furthermore, we believe that ENDOR spectroscopy has a high potential for studying the mechanisms of aggregation and disaggregation of conglomerates of vanadyl porphyrins. The similarity of the experimental ENDOR spectra for two very different crude oil samples reported in our work seems to be due to universal (at least for some types of oils) supermolecular structures. Whether they belong to the asphaltenes aggregates or to other high-molecular weight phases is an open question. Until recently a role of metalloporphyrins in the aggregation of the high-molecular weight phases was discounted or even denied. Quite fresh experimental and theoretical studies show that vanadyls can be of crucial importance for formation of the nano-, micro- and macro aggregates [18, 37, 38]. We are planning to address this issue in our next publications.

**Acknowledgments**

The authors devote this work to EPR expert Dr. I.N. Kurkin (Kazan) on the occasion of his 75$^{th}$ anniversary. We are thankful to Dr. Mikhail Varfolomeev (Kazan) for the part of the provided samples and their characterization. The work is financially supported by the Program of the competitive growth of Kazan Federal University among the World scientific centers "5-100".

**Supporting Information**

The manuscript contains supporting information. This information is available free of charge via the Internet at http://pubs.acs.org/.

**Figure S1** shows the relative contributions of EPR transitions for $VO^{2+}$ paramagnetic complexes with different $m_I$ at $\nu = 94$ GHz in the magnetic fields perpendicular ($B_{1'}$) and parallel ($B_{2'}$) to the direction of the magnetic field $B_0$ for $m_I = 3/2$. **Appendix S1** is an EasySpin script for $^1$H ENDOR simulation. **Appendix S2** is a list of XYZ coordinates of the DFT optimized vanadyl porphyrin models considered in the paper.

**Supporting Information for the Manuscript:**

**In-situ identification of various structural features of vanadyl porphyrins in crude oil by high-field (3.4 T) ENDOR spectroscopy combined with DFT calculations**

*Timur Biktagirov[a,b], Marat Gafurov[a,*], Georgy Mamin[a], Irina Gracheva[a],*

*Andrey Galukhin[b,c], Sergei Orlinskii[a]*

[a] *Institute of Physics, Kazan Federal University, 420008 Kazan, Russia*

[b] *Institute of Geology and Petroleum Technology, Kazan Federal University, 420008 Kazan, Russia*

[c] *Institute of Chemistry, Kazan Federal University, 420008 Kazan, Russia*

[†]to whom correspondence should be addressed
e-mail: marat.gafurov@kpfu.ru

**Fig. S1.** Relative contributions of EPR transitions for VO$^{2+}$ paramagnetic complexes with different $m_I$ at $\nu$ = 94 GHz in the magnetic fields perpendicular ($B_{1'}$) and parallel ($B_{2'}$) to the direction of the magnetic field $B_0$ for $m_I$ = 3/2 as a function of angle between the direcrtion of $B_0$ and the directions of the parallel component of g ($g_\parallel$) . The parameters of calculation are the same as for Figure 2 of the Manuscript.

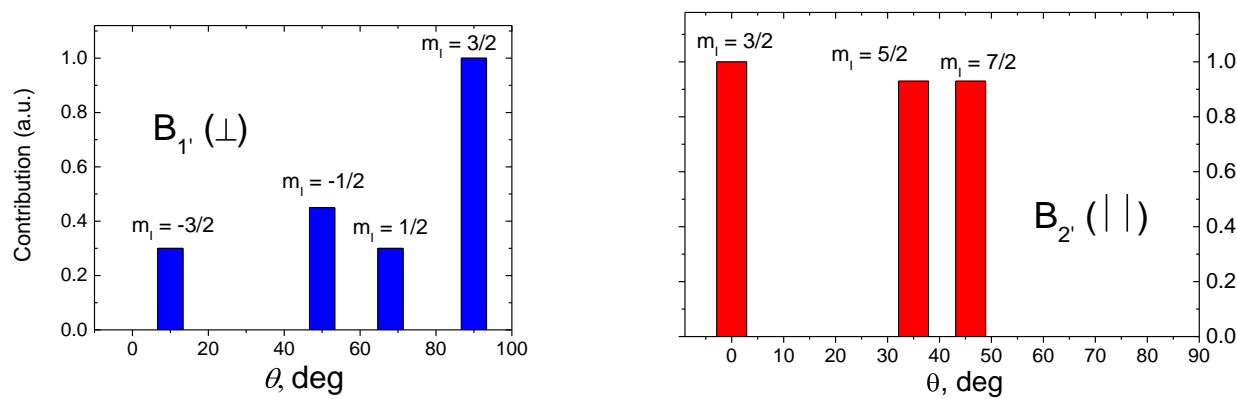

**Appendix S1.** Easyspin script (for further details, see Easyspin online manual)

%This is the example of ENDOR simulation for the H1 proton. Hyperfine coupling parameters for vanadium and g-tensor are determined by the EPR spectrum simulation.

```
clear, clf

% Spin system:
%nuclei
Sys.Nucs = '51V, 1H';
%principle components of the hyperfine tensors for 51V (variable AV) and 1H (variable AH) in
AV = [156.9, 156.9, 470.8];
AH = [-0.2, -0.2, 2.4];
Sys.A=[AV; AH];
%and the corresponding Euler angles in radians
AFrameV=[0.0, 0.0, 0.0];
AFrameH = [1.6, 1.5, -3.1];
Sys.AFrame = [AFrameV; AFrameH];
%g-factor with its orientation
Sys.g = [1.9845, 1.9845, 1.9640];
Sys.gFrame=[0.0, 0.0, 0.0];
%ENDOR linewidth
Sys.lwEndor = 0.4;

% Experiment settings:
%microwave frequency
Exp.mwFreq = 94.60;
%excitation width
```

```
Exp.ExciteWidth = 60;
%radiofrequency sweeping range
Exp.Range = [138 148];
%number of points in the spectrum
Exp.nPoints = 1001;

%Simulation options:
Opt.Method = 'perturb1';
Opt.nKnots = 90;

%Magnetic field values:
Fields = [3396, 3449];

%Simulation:
for  iField = 1:numel(Fields)
  Exp.Field = Fields(iField);
  [freq,spectra(iField,:)] = salt(Sys,Exp, Opt);
end
```

**Appendix S2.** XYZ coordinates (in Å) of the DFT optimized vanadyl porphyrin models considered in the paper

**VO**

| | | | |
|---|---|---|---|
| V | -0.662191129 | -0.012516335 | 0.553973969 |
| N | 0.873165528 | 1.294644668 | 0.034936064 |
| N | -2.206465849 | -1.318445773 | 0.058614855 |
| C | -2.127041246 | -2.688855950 | 0.014200496 |
| N | 0.639819926 | -1.551771424 | 0.033587416 |
| C | -3.446778988 | -3.266569340 | -0.039700761 |
| C | -3.545962551 | -1.017215541 | 0.028524171 |
| C | -4.330889235 | -2.224864083 | -0.030923466 |
| C | 0.338276964 | -2.890909270 | -0.007366428 |
| C | -0.946759256 | -3.427964776 | -0.003286902 |
| C | 2.009870877 | -1.471908733 | -0.020161173 |
| C | 2.587061939 | -2.791055170 | -0.091468059 |
| C | 1.545399389 | -3.675218793 | -0.083300071 |
| C | 2.211932002 | 0.993511112 | -0.019087637 |
| C | 2.748907696 | -0.291507119 | -0.031616235 |
| C | 2.995723121 | 2.201308272 | -0.089496344 |
| C | 0.793002603 | 2.665150012 | -0.004760561 |
| C | -0.387391722 | 3.404259672 | -0.000060990 |
| C | 2.111637105 | 3.243024224 | -0.079976005 |
| N | -1.973145294 | 1.527950107 | 0.059957687 |
| C | -1.672319738 | 2.867177503 | 0.016768154 |
| C | -3.343913297 | 1.448191393 | 0.029604417 |
| C | -4.083058610 | 0.267813448 | 0.028262367 |
| C | -3.922218626 | 2.767479352 | -0.028873294 |
| C | -2.880554608 | 3.651647186 | -0.036533584 |
| O | -0.648512213 | -0.014384970 | 2.115366482 |
| H | 2.320591911 | 4.305255186 | -0.128968116 |
| H | 4.077371223 | 2.235304813 | -0.147922548 |
| H | -5.413415592 | -2.258765483 | -0.070274901 |
| H | -3.656560263 | -4.328674743 | -0.087800720 |
| H | -5.165442797 | 0.356877971 | 0.002021336 |
| H | -4.984760749 | 2.976887636 | -0.068252652 |
| H | -2.914919422 | 4.733842409 | -0.083680003 |
| H | -0.298657292 | 4.486420446 | -0.035025302 |
| H | 3.830665718 | -0.380508511 | -0.076745666 |
| H | 3.648754494 | -3.000360655 | -0.149676799 |
| H | 1.578917666 | -4.757301687 | -0.133463048 |
| H | -1.036109682 | -4.510037053 | -0.039406145 |

**VOEtio**

| | | | |
|---|---|---|---|
| V | -0.625654132 | -0.004996967 | 0.205544860 |
| N | 0.898490589 | 1.311420983 | -0.370340741 |
| N | -2.156458393 | -1.368917239 | -0.232472447 |
| C | -2.045072659 | -2.729075953 | -0.199587361 |
| N | 0.715015457 | -1.553617272 | -0.237994438 |
| C | -3.355474275 | -3.343901472 | -0.220237886 |
| C | -3.491908349 | -1.090790032 | -0.276409642 |
| C | -4.264712918 | -2.317114138 | -0.272023633 |
| C | 0.432241089 | -2.888304749 | -0.202079204 |
| C | -0.846859734 | -3.435830078 | -0.173890696 |
| C | 2.074978951 | -1.447181203 | -0.287684863 |
| C | 2.685740219 | -2.761646185 | -0.281215059 |
| C | 1.653708348 | -3.664906681 | -0.224529847 |
| C | 2.233273522 | 1.029388084 | -0.404295253 |
| C | 2.780395316 | -0.249794117 | -0.362086393 |
| C | 3.005755077 | 2.250271139 | -0.520161173 |
| C | 0.786874845 | 2.668795649 | -0.462258057 |
| C | -0.410865528 | 3.376606937 | -0.489920054 |
| C | 2.096776230 | 3.278260962 | -0.554343524 |
| N | -1.970033033 | 1.498783270 | -0.362499447 |
| C | -1.691063870 | 2.831891205 | -0.450932992 |
| C | -3.329949346 | 1.384918685 | -0.390118227 |
| C | -4.037782146 | 0.187730542 | -0.344220995 |
| C | -3.944107045 | 2.691542646 | -0.496525475 |
| C | -2.916726071 | 3.600843548 | -0.535665452 |
| O | -0.618100877 | 0.066077883 | 1.779360740 |
| C | 2.350358204 | 4.744972110 | -0.661817544 |
| C | 4.500044709 | 2.331400972 | -0.546767955 |
| C | -5.759028311 | -2.402410038 | -0.274450624 |
| C | -3.610356898 | -4.813808952 | -0.187464253 |
| H | -5.127504974 | 0.261426284 | -0.381800582 |
| C | -5.414284238 | 2.942897102 | -0.546670272 |
| C | -3.002781063 | 5.093546676 | -0.602756053 |
| H | -0.335440298 | 4.463723948 | -0.571051369 |
| H | 3.869619673 | -0.322421728 | -0.409407840 |
| C | 4.157608881 | -3.033920920 | -0.287671153 |
| C | 1.719870100 | -5.154988769 | -0.190056594 |
| H | -0.918357217 | -4.526149193 | -0.147780294 |
| H | 2.016907979 | 5.282740472 | 0.244974296 |
| H | 1.813698187 | 5.190892811 | -1.518787609 |
| H | 3.422780493 | 4.959983529 | -0.795303649 |
| H | -3.085432811 | -5.339027914 | -1.005906816 |
| H | -4.684556026 | -5.039079891 | -0.285410528 |
| H | -3.264672163 | -5.264802431 | 0.761033562 |
| H | 1.337578370 | -5.557146683 | 0.766266712 |
| H | 2.754586043 | -5.514413131 | -0.307154128 |
| H | 1.117470188 | -5.609490184 | -0.997105033 |
| H | -5.635571991 | 4.011962876 | -0.696070070 |
| H | -5.911995363 | 2.631174795 | 0.390146933 |
| H | -5.896531608 | 2.386005076 | -1.370421356 |
| C | 5.139390559 | 2.245657994 | 0.841191511 |
| H | 4.804420747 | 3.276978154 | -1.028702297 |
| H | 4.906329045 | 1.527901347 | -1.187580422 |
| H | 4.787242722 | 3.065828765 | 1.490482958 |
| H | 6.239919729 | 2.309945556 | 0.773184496 |
| H | 4.882343527 | 1.296206391 | 1.342932520 |
| C | 4.806366105 | -2.895425787 | 1.091893999 |
| H | 4.660467352 | -2.354412175 | -0.999754751 |
| H | 4.337562487 | -4.052754004 | -0.672822756 |
| H | 4.674753975 | -1.876364111 | 1.495957639 |
| H | 5.889179629 | -3.106761466 | 1.041718345 |
| H | 4.352521004 | -3.596466783 | 1.813048425 |
| C | -2.894172604 | 5.770541743 | 0.765619210 |
| H | -3.958219394 | 5.382068202 | -1.075088183 |
| H | -2.212822648 | 5.483857043 | -1.269538969 |
| H | -3.704513990 | 5.435971567 | 1.436281096 |
| H | -2.959452007 | 6.868939380 | 0.669369152 |
| H | -1.937210594 | 5.525487869 | 1.258754815 |
| C | -6.382235262 | -2.240548738 | 1.114045357 |
| H | -6.066463446 | -3.374316080 | -0.698349845 |
| H | -6.175294095 | -1.637897693 | -0.954981263 |
| H | -6.022815405 | -3.025404873 | 1.801817172 |
| H | -7.483705310 | -2.306460138 | 1.063141022 |
| H | -6.116713258 | -1.266224425 | 1.559660245 |

**VODPEP**

| | | | |
|---|---:|---:|---:|
| V | -0.643745191 | 0.002943068 | 0.252819570 |
| N | 0.780094898 | 1.316079750 | -0.368160212 |
| N | -2.127681979 | -1.342300772 | -0.234981596 |
| C | -2.006234434 | -2.706459338 | -0.208655954 |
| N | 0.767158701 | -1.515034480 | -0.202881483 |
| C | -3.309201596 | -3.329527742 | -0.251225607 |
| C | -3.468312213 | -1.080702618 | -0.284012148 |
| C | -4.225143554 | -2.313496065 | -0.301166084 |
| C | 0.472111191 | -2.847521156 | -0.166966425 |
| C | -0.807390617 | -3.396636434 | -0.161962916 |
| C | 2.144191806 | -1.434820499 | -0.252670604 |
| C | 2.726244233 | -2.756247098 | -0.229444821 |
| C | 1.676895202 | -3.637725182 | -0.175929376 |
| C | 2.087053137 | 0.993581121 | -0.407231415 |
| C | 2.786861435 | -0.200970225 | -0.358984797 |
| C | 2.924395319 | 2.132219429 | -0.536419857 |
| C | 0.755457249 | 2.687307279 | -0.467102855 |
| C | -0.429695067 | 3.410678504 | -0.472800957 |
| C | 2.100780783 | 3.226264696 | -0.578864210 |
| N | -1.965532867 | 1.505486317 | -0.320082870 |
| C | -1.698466051 | 2.848027793 | -0.412629513 |
| C | -3.327995994 | 1.388991677 | -0.353662513 |
| C | -4.026508623 | 0.190386049 | -0.328911617 |
| C | -3.948610859 | 2.690427985 | -0.456231261 |
| C | -2.930652988 | 3.603789754 | -0.492950619 |
| O | -0.635175011 | 0.056785010 | 1.808391857 |
| C | 2.443806986 | 4.673183082 | -0.707963201 |
| C | -5.717950790 | -2.408089052 | -0.324945602 |
| C | -3.550486951 | -4.801928570 | -0.238465197 |
| H | -5.109850735 | 0.254328771 | -0.368557532 |
| C | -5.420525335 | 2.929430654 | -0.511252613 |
| C | -3.022449021 | 5.095187436 | -0.564837679 |
| H | -0.361052359 | 4.491229186 | -0.556821561 |
| C | 4.183235033 | -3.103443297 | -0.206011254 |
| C | 1.722444037 | -5.128866948 | -0.132897695 |
| H | -0.870577497 | -4.480861005 | -0.139411619 |
| H | 2.126540001 | 5.245457969 | 0.173417972 |
| H | 1.960138700 | 5.130009710 | -1.580614439 |
| H | 3.523688942 | 4.808509360 | -0.819065945 |
| H | -3.008807476 | -5.307791318 | -1.047375967 |
| H | -4.612798289 | -5.033855425 | -0.357287441 |
| H | -3.219383655 | -5.255264420 | 0.704799540 |
| H | 1.336771729 | -5.517477025 | 0.818419198 |
| H | 2.744670704 | -5.500561201 | -0.246447872 |
| H | 1.119242572 | -5.575639704 | -0.932893593 |
| H | -5.648332751 | 3.989094776 | -0.657971710 |
| H | -5.914305227 | 2.612763824 | 0.416545870 |
| H | -5.889976681 | 2.373590003 | -1.332588486 |
| C | 4.761772695 | -3.129034454 | 1.212206587 |
| H | 4.756144836 | -2.405885612 | -0.825203229 |
| H | 4.323577520 | -4.090347271 | -0.664718345 |
| H | 4.638245027 | -2.160978655 | 1.710354797 |
| H | 5.830125133 | -3.372718236 | 1.196937560 |
| H | 4.248117890 | -3.877331937 | 1.824825072 |
| C | -2.848416844 | 5.770369545 | 0.799123774 |
| H | -3.992872109 | 5.380697095 | -0.988214496 |
| H | -2.268337692 | 5.481467068 | -1.263279411 |
| H | -3.623490234 | 5.438819891 | 1.498245507 |
| H | -2.912062191 | 6.860715375 | 0.707922451 |
| H | -1.878471555 | 5.519075896 | 1.241440887 |
| C | -6.350879369 | -2.220205252 | 1.057205908 |
| H | -6.014842616 | -3.383354637 | -0.728965195 |
| H | -6.126491208 | -1.661997060 | -1.019200859 |
| H | -5.997722580 | -2.987784275 | 1.754033489 |
| H | -7.443514520 | -2.284971864 | 1.000385874 |
| H | -6.086512136 | -1.245686548 | 1.481357586 |
| C | 4.262810417 | 0.136711496 | -0.491122233 |
| C | 4.354092298 | 1.696323217 | -0.563190408 |
| H | 4.687001824 | -0.309538596 | -1.399775501 |
| H | 4.849735167 | -0.254421658 | 0.348551108 |
| H | 4.879655954 | 2.021721344 | -1.469676775 |
| H | 4.921255445 | 2.096421499 | 0.286520960 |

**VOBenzo**

| | | | |
|---|---|---|---|
| V | -0.619785098 | -0.021132011 | 0.232636924 |
| N | 0.885576287 | 1.288879753 | -0.352436176 |
| N | -2.134508636 | -1.374669600 | -0.217601685 |
| C | -2.026004092 | -2.741986843 | -0.191539039 |
| N | 0.711130564 | -1.551212628 | -0.229524972 |
| C | -3.331303146 | -3.357221416 | -0.209680581 |
| C | -3.474609897 | -1.110434114 | -0.250595367 |
| C | -4.241396811 | -2.335556648 | -0.249505771 |
| C | 0.438882053 | -2.892389878 | -0.202801154 |
| C | -0.833998687 | -3.444980389 | -0.174072179 |
| C | 2.074924696 | -1.450352245 | -0.285381444 |
| C | 2.685340008 | -2.758726841 | -0.293330303 |
| C | 1.657048839 | -3.662542151 | -0.239443921 |
| C | 2.228648452 | 1.014323460 | -0.390064023 |
| C | 2.778881803 | -0.255769146 | -0.353923019 |
| C | 2.996202568 | 2.233689882 | -0.499761069 |
| C | 0.783279138 | 2.649381683 | -0.433522966 |
| C | -0.409004424 | 3.360954210 | -0.453819431 |
| C | 2.088787688 | 3.258322254 | -0.525135397 |
| N | -1.974633433 | 1.485598827 | -0.326570473 |
| C | -1.680118133 | 2.819783888 | -0.419818623 |
| C | -3.337297283 | 1.353969020 | -0.354889612 |
| C | -4.032127835 | 0.160403025 | -0.307352872 |
| C | -3.955789475 | 2.660289108 | -0.467803077 |
| C | -2.905401984 | 3.589495824 | -0.508368918 |
| O | -0.612343032 | 0.047885520 | 1.784963151 |
| C | 2.339501783 | 4.725869821 | -0.626624614 |
| C | 4.489616527 | 2.313432773 | -0.533349345 |
| C | -5.734999886 | -2.420114969 | -0.250388118 |
| C | -3.581829528 | -4.827987276 | -0.187220350 |
| H | -5.115706529 | 0.224806873 | -0.339784377 |
| H | -0.335656600 | 4.441746457 | -0.527712821 |
| H | 3.861384259 | -0.325615097 | -0.403000507 |
| C | 4.157274716 | -3.026049691 | -0.315672144 |
| C | 1.726560326 | -5.152923056 | -0.222749022 |
| H | -0.901490369 | -4.528831674 | -0.154856575 |
| H | 2.010698368 | 5.254083417 | 0.277698947 |
| H | 1.803127777 | 5.170411458 | -1.474340139 |
| H | 3.403540385 | 4.940251997 | -0.761711333 |
| H | -3.058596000 | -5.340907882 | -1.003807111 |
| H | -4.647672104 | -5.052640577 | -0.286618708 |
| H | -3.237811591 | -5.280282550 | 0.751855830 |
| H | 1.355999050 | -5.563979251 | 0.725104796 |
| H | 2.754045316 | -5.504442876 | -0.351558315 |
| H | 1.123625505 | -5.595217270 | -1.025396331 |
| C | 5.125882003 | 2.203191146 | 0.855690841 |
| H | 4.793131008 | 3.260613826 | -0.995431205 |
| H | 4.890478727 | 1.524389031 | -1.183129441 |
| H | 4.779886060 | 3.012467064 | 1.507564161 |
| H | 6.218483748 | 2.257881708 | 0.791384678 |
| H | 4.858087976 | 1.256860327 | 1.338284280 |
| C | 4.812189088 | -2.883502147 | 1.061837242 |
| H | 4.648689956 | -2.348143871 | -1.026207231 |
| H | 4.337867723 | -4.037766351 | -0.698439480 |
| H | 4.669469690 | -1.874287432 | 1.463550179 |
| H | 5.888689676 | -3.080821662 | 1.006310931 |
| H | 4.371197200 | -3.586128989 | 1.776443137 |
| C | -6.346001742 | -2.224085533 | 1.140487205 |
| H | -6.043577824 | -3.394738218 | -0.646930955 |
| H | -6.149858760 | -1.674494878 | -0.941405236 |
| H | -5.986935456 | -2.991461202 | 1.834485573 |
| H | -7.439679030 | -2.283101882 | 1.100132985 |
| H | -6.070271911 | -1.249844699 | 1.557969708 |
| C | -3.164369472 | 4.957397917 | -0.622577761 |
| C | -5.285967734 | 3.080266013 | -0.539943815 |
| C | -4.486642846 | 5.370296784 | -0.690918090 |
| H | -2.358596710 | 5.686004109 | -0.656255826 |
| C | -5.538619634 | 4.439386066 | -0.649902207 |
| H | -6.107553329 | 2.369057337 | -0.509526437 |
| H | -4.716160920 | 6.428954476 | -0.778039568 |
| H | -6.564549021 | 4.793147884 | -0.705141438 |

--------------------------------------------------------------------------------